\documentclass[a4paper,12pt]{article}
\usepackage[pctex32]{graphics}
\usepackage{amssymb,amsmath}

\begin{document}

\topmargin 0pt \oddsidemargin 0mm
\newcommand{\be}{\begin{equation}}
\newcommand{\ee}{\end{equation}}
\newcommand{\ba}{\begin{eqnarray}}
\newcommand{\ea}{\end{eqnarray}}
\newcommand{\fr}{\frac}

\begin{titlepage}

\vspace{5mm}
\begin{center}

{\Large \bf Phase transition  between non-extremal and extremal
Reissner-Nordstr\"om  black holes}

\vspace{12mm}

{\large   Yun Soo Myung\footnote{e-mail
 address: ysmyung@inje.ac.kr}}
 \\
\vspace{10mm} {\em Institute of Mathematical Science and School of
Computer Aided Science, \\Inje University, Gimhae 621-749, Korea
\\}

\end{center}

\vspace{5mm}

\centerline{{\bf{Abstract}}} We discuss the phase transition
between non-extremal and extremal Reissner-Nordstr\"om black
holes. This transition is considered as the
 $T \to 0$ limit of the transition between the non-extremal and near-extremal black holes.
  We show that an
evaporating process from non-extremal black hole to extremal one
is possible to occur, but  its reverse process  is not possible to
occur because of the presence of the maximum temperature.
 Furthermore, it is shown that the Hawking-Page phase transition
 between small and large black holes  unlikely occurs in the AdS Reissner-Nordstr\"om black holes.

\vspace{5mm}

\vspace{3mm}


\end{titlepage}

\renewcommand{\thefootnote}{\arabic{footnote}}
\setcounter{footnote}{0} \setcounter{page}{2}

\section{ Introduction}
Since the pioneering work of Hawking-Page on the phase transition
between thermal AdS and AdS black hole in four
dimensions~\cite{HP}, the research of the black hole
thermodynamics has greatly improved. Especially, the phase
transition of  the AdS black hole in five dimensions inspired by
string theory has generated renewed attention  because it relates
to confining-deconfining phase transition on the gauge theory side
through the AdS/CFT duality~\cite{Witt}. It was extended to study
the transition between AdS soliton and AdS black hole with Ricci
flat horizon~\cite{HM,SSW}.

In addition to the usual transition between thermal AdS and AdS
large black hole, it was suggested that there may exist a
different phase transition between small and large black holes
(HP1) in the AdS Reissner-Nordstr\"om (AdSRN) black holes for
fixed charge $Q<Q_c$~\cite{CEJM,DMMS1} and AdS Gauss-Bonnet black
holes~\cite{DMMS2}.

 In the conventional Hawking-Page phase transition (HP2), one generally starts with
thermal radiation in AdS space. An unstable small black hole
appears with negative heat capacity. The heat capacity changes
from negative infinity to positive infinity at the minimum
temperature $T_0$. Finally, the large black hole with positive
heat capacity comes out as a stable object. There is a change of
the dominance at the critical temperature $T_1$: from thermal
radiation  to black hole~\cite{HP}.

In the HP1, we start with a stable small stable black hole with
positive heat capacity (SBH)  because the thermal AdS is not an
admissible phase for the AdSRN black holes~\cite{DMMS1}. The heat
capacity changes from positive infinity to negative infinity at
the maximum temperature $T_m$. Then, an intermediate black hole
 with negative heat capacity (IBH) comes out. There is a change of
the dominance at the critical temperature $T=T_2$: from SBH to
large black hole  with positive heat capacity (LBH). The HP1 can
be described by two processes: SBH $\to$ IBH $\to$ LBH. However,
it seems that the first process of SBH $\to$ IBH is not permitted
because there exists a temperature barrier between SBH and IBH.

In order to investigate  the first process of SBH $\to$ IBH, we
have to study thermodynamics of the RN black hole~\cite{MKP}. In
general, non-extremal RN black holes  form two parameters of mass
$M$ and charge $Q$ with the inequality of $M>|Q|$. Here NOBH
denotes non-extremal Reissner-Norstr\"om black hole with
$M>M_m=2Q/\sqrt{3}$.  On the other hand, extremal RN black holes
(EBH) form one parameter with $M=|Q|$. Their surface gravity
vanishes and thus their Hawking temperature is zero. Therefore,
for fixed-charge ensemble, the EBH can be the stable endpoint of
the evaporation of NOBH. Recently, it was reported that there is
no real discontinuity between EBH and NOBH, if one takes into
account the flux due to the Hawking radiation
properly~\cite{BFFP}.

In this work, we first check  whether the first process of HP1
occurs or not by studying the phase transition  between NOBH and
near-extremal black hole (NEBH) with $Q<M<M_m$. In order to
investigate this transition, we review thermodynamics of
Reissner-Nordstr\"om black hole thoroughly. We study the off-shell
(non-equilibrium) process of the growth of the RN black hole using
the off-shell
 free energy and off-shell $\beta$-function. It
turns out that the process of SBH $\to$ IBH is not allowed  but
its reverse process of SBH $\leftarrow$ IBH is permitted as an
evaporating process. Hence, it is shown that the HP1 of
SBH$\to$IBH$\to$LBH unlikely occurs in the AdSRN black holes.

\section{Thermodynamics of RN black hole}

We consider the RN black hole whose metric is given by
\begin{equation} \label{MF}
ds^2_{RN}=-U(r)dt^2+U^{-1}(r)dr^2+r^2d\theta^2+r^2\sin^2\theta
d\varphi^2
\end{equation}
with $U(r)=1-2M/r+Q^2/r^2$. Here, $M$ and $Q$ are the mass and the
electric charge of the RN black hole, respectively. Then, the
inner ($r_-$) and the outer ($r_+$) horizons are obtained as
$r_\pm=M\pm\sqrt{M^2-Q^2}$, which satisfy $U(r_\pm)=0$. For $M=Q$,
we have an extremal RN black hole at $r_+=Q$. In this work we
consider the case of fixed charge $Q$ for simplicity~\cite{CEJM}.
The other case of the fixed potential $\Phi=Q/r_+$ will have
parallel with the fixed charge case.

For the RN black hole, the relevant thermodynamic quantities are
given by the ADM mass $M$, Bekenstein-Hawking entropy $S_{BH}$ and
Hawking temperature $T_H$:
\begin{eqnarray} \label{as}
M(r_+,Q)&=&\frac{r_+}{2}\Big(1+\frac{Q^2}{r_+^2}\Big),~S_{BH}(r_+)= \pi r_+^2,\\
T_{H}(r_+,Q)&=& \frac{1}{4\pi}\Big(
\frac{1}{r_+}-\frac{Q^2}{r_+^3}\Big). \label{at}
\end{eqnarray}
Then, using the Eqs. (\ref{as}) and (\ref{at}), the heat capacity
$C=(dM/dT_{H})_Q$ for fixed charge and Helmholtz free energy $F$
are obtained to be
\begin{eqnarray}\label{ac}
C(r_+,Q)&=& -2\pi r_+^2 \Big(\frac{r_+^2-Q^2}{r_+^2-3Q^2}\Big), \\
\label{af} F(r_+,Q)&=&E-T_{H}S_{BH}=\frac{r_+^2+3Q^2}{4r_+}-Q
\end{eqnarray}
with $E=M-Q$. In this case, $E$ measures the energy above the
ground state. We note  that one has to use the extremal black hole
as background for fixed charge ensemble~\cite{CEJM}.
\begin{figure*}[t!]
   \centering
   \includegraphics{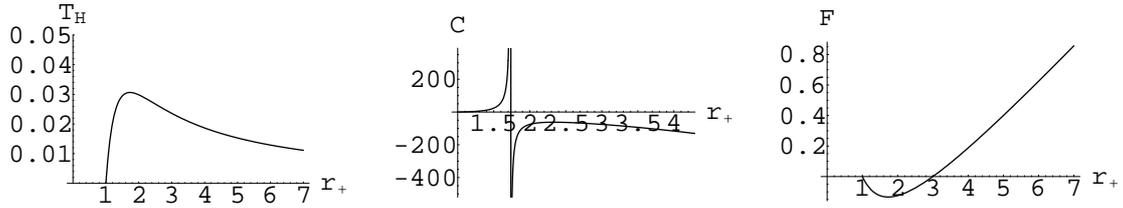}
\caption{Thermodynamic quantities of RN black hole as function of
horizon radius $r_+$ with fixed $Q=1$: temperature $T_H$, heat
capacity $C$, and free energy $F$. } \label{fig1}
\end{figure*}

As is shown in Fig. 1, the features of thermodynamic quantities
are as follows: First of all,  the whole region splits into the
near-horizon phase of $Q<r_+<r_m$ with $r_m=\sqrt{3}Q$ and
Schwarzschild phase of $r_+>r_m$.  i) The temperature is zero
($T_H=0$)  at $r_+=Q$, maximum $T_m=\frac{1}{6\sqrt{3} \pi Q}$ at
$r_+=r_m$, and for $r_+>r_m$, it shows the Schwarzschild behavior.
ii) The heat capacity $C$ determines the local stability of
thermodynamic system. For $Q<r_+<r_m$, it is locally stable
because of $C>0$, while for $r_+>r_m$, it is locally  unstable
($C<0$) as is shown in the Schwarzschild case. Here, we observe
that  $C=0$ at $r_+=Q$ and importantly, it blows up at $r_+=r_m$.
iii) The free energy is an important quantity to determine where
the presumed phase transition occurs. The free energy  is negative
 for the near-horizon phase and it is increasing
 for the Schwarzschild phase. It takes the minimum
value of $F_m=-\frac{Q(2-\sqrt{3})}{2}$ at $r_+=r_m$ and $F=0$ at
$r_+=Q,~3Q$. We identify the thermal state of the EBH  by the
condition of $T_H=0,~C=0,~F=0$ with the non-zero
Bekenstein-Hawking entropy $S_{BH}=\pi Q^2$~\cite{MKP}.
Furthermore, the RN black hole  is split into NEBH being in the
region of $Q<r_+<r_m$ and NOBH in the region of $r_+>r_m$. In
connection with HP1, we have a correspondence of NEBH
$\leftrightarrow$ SBH and NOBH $\leftrightarrow$ IBH.

\section{Off-shell (non-equilibrium) process}
\begin{figure*}[t!]
   \centering
   \includegraphics{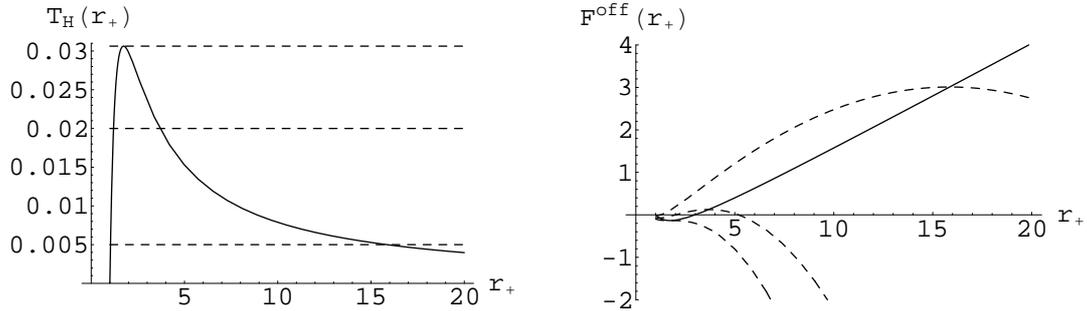}
\caption{Graphs for temperature and free energy expressed as
functions of $r_+$. In the left panel, the solid curve represents
the Hawking temperature $T_H(r_+,Q=1)$, while three dashed lines
denote $T=T_m=0.03,~0.02,~0.005$ from top to bottom. In the right
panel, the solid curve represent the on-shell free energy
$F(r_+,Q=1)$, while three dashed curves denote the off-shell free
energy $F^{off}(r_+,Q=1,T)$ for $T=0.005,0.02,T_m$ from top to
bottom.} \label{fig2}
\end{figure*}
\begin{figure*}[t!]
   \centering
   \includegraphics{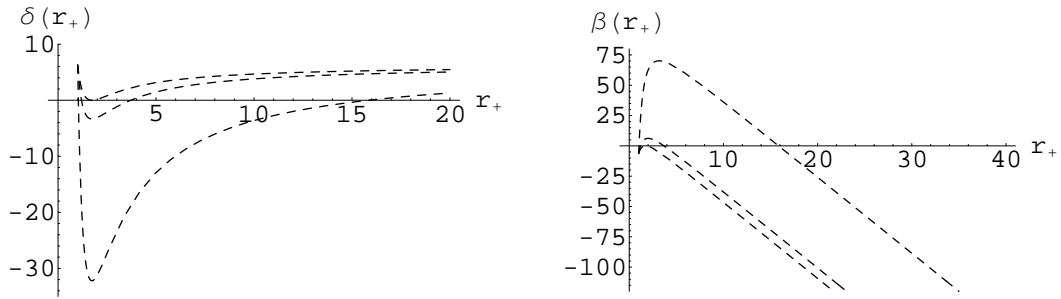}
\caption{Graphs for deficit angle and $\beta$-function. Three
dashed curves denote $\delta(r_+,Q=1,T)$ for $T=T_m, 0.02,0.005$
from top to bottom. In the right panel, three dashed curves denote
$\beta(r_+,Q=1,T)$ for $T=0.005,0.02,T_m$ from top to bottom. }
\label{fig3}
\end{figure*}
We use the off-shell free energy to study the growth of a black
hole~\cite{FFZ}. Also, the off-shell $\beta$-function is
introduced to measure the mass of a conical singularity at the
event horizon~\cite{off1,off2,off3}. In order to explore a
possible phase transition, we consider  the off-shell free energy,
\begin{equation} \label{offfree}
F^{off}(r_+,Q,T)=E-TS_{BH}
\end{equation}
with the external temperature $T$. For fixed $Q$, it corresponds
to the Landau function for describing the phase transition with
the order parameter $r_+$~\cite{DMMS1}. We note that $\partial
F^{off}/\partial r_+=0 \to T=T_H$. Plugging this into
Eq.(\ref{offfree}) leads to the on-shell free energy $F(r_+,Q)$ in
Eq.(\ref{af}). In this sense, we call $F^{off}$ the off-shell free
energy~\cite{myung11,myung12}. In other words, the first law of
thermodynamics of $dE=TdS_{BH}$ holds for the on-shell but it does
not hold for any  off-shell process.

We start with observing the temperature in Fig. 2. An important
point is that there exists the maximum temperature $T=T_m=0.3$
near the extremal black hole. The presence of this temperature
makes the thermodynamic process different when comparing with the
Hawking-Page phase transition (HP2) between thermal AdS and black
hole~\cite{myung2}. Here we choose three external temperatures to
study a possible phase transition between NEBH  and NOBH. These
are $T=T_m,~0.02,~0.005$. It is noted that in the limit of $T\to
0$, one may include  the phase transition between EBH and NOBH. In
this case, the starting point approaches the extremal black hole
($r_i \to Q$), while the ending point goes to infinity ($r_f \to
\infty$). For these temperatures, the corresponding graphs of
off-shell free energy are shown in Fig. 2. Considering the
condition of $T=T_H$, we find their equilibrium points as the
initial and final  points. These are given by
\begin{equation}
  r_i=\frac{1}{24 \pi T}\Big[2+ 4
\cos\Big(\frac{\alpha}{3}-\frac{2\pi}{3}\Big)\Big],~r_f=\frac{1}{24
\pi T}\Big[2+ 4 \cos\frac{\alpha}{3}\Big],
\end{equation}
 where $\cos
\alpha=1-2\frac{T^2}{T_m^2}$ with $2\frac{T}{T_m} < \alpha \le
\pi$. In particular, for  $T \ll T_m~(\alpha \to
2\frac{T^2}{T^2_m}\simeq 0)$, the initial point  is located at
$r_i \simeq r_e$,
 while the final point  is at $r_f\simeq \frac{1}{4\pi T}$.
In case of $T=T_m(\alpha = \pi)$, one has a degenerate point  at
$r_i=r_f=r_m$, which corresponds to the maximum point.

 As is expected from the temperature graph,
for $T>T_m$, there is no equilibrium point where the on-shell
curve of free energy meets the off-shell curve ($F=F^{off}$). For
the maximum temperature $T=T_m$, there is a degenerate equilibrium
point at $r_+=r_m$, where the minimum on-shell free energy
$F_m=F^{off}$ appears and the heat capacity blows up. In  case of
$T=0.02$, we have two equilibrium points $r_i=1.2$ and $r_f=3.69$:
$r_i$ is a globally stable point  because of $C>0,F<0$ while $r_f$
is a globally unstable point because of $C<0,F>0$. For $T=0.005$,
the situation is qualitatively similar to the case of $T=0.02$,
but the difference is  $r_i=1.03$ and $r_f=15.85$. Hence, the
transition from NOBH  to NEBH is likely allowed as an evaporating
process~\cite{Pavon}, while its reverse process  is unlikely
allowed because of the presence of the maximum temperature
$T=T_m$.

In order to investigate the off-shell process explicitly,  we
consider the off-shell parameter $\alpha$ and the deficit angle
$\delta$ as
\begin{equation}  \label{delta}
\alpha(r_+,Q,T)=\frac{T_{H}}{T},~~\delta(r_+,Q,T)=2 \pi(1-\alpha).
\end{equation}
$\alpha$ is zero at the extremal point of $r_+=Q$ and it is one at
the equilibrium point $T=T_H$. On the other hand, the range of
deficit angle is given by $0\le \delta \le 2\pi$. $\delta$ has the
maximum value of $2\pi$ at the extremal point and it is zero at
$T=T_H$. This implies that the extremal configuration at $r_+=Q$
has the narrowest cone of the shape ($\prec$) near the horizon,
while its geometry at $T=T_H$ is a contractible manifold
($\subset$) without conical singularity. For any off-shell process
of  the growth of black hole, we must have $0<\delta<2\pi$ and a
conical singularity of the shape ($<$) near the
horizon~\cite{FFZ,off1,off2,off3,myung11,myung12}. However, as is
shown in Fig. 3, there exist negative deficit angles for $T<T_m$,
while the deficit angle is always positive for $T>T_m$.
Explicitly, we find negative deficit angles for $r_i<r_+<r_f$.
Hence we stress that these are not properly defined  thermodynamic
processes  of growing black hole by absorbing radiation in the
heat reservoir.

Furthermore, the off-shell process could be  described by the
off-shell $\beta$-function defined with the off-shell Euclidean
action $I^{off}=F^{off}/T$ as
\begin{equation}\label{beta}
\beta(r_+,Q,T)\propto \frac{\partial I^{off}}{\partial r_+} =-
r_+~ \delta(r_+,Q,T)
\end{equation}
which measures the deficit angle $\delta$ directly. The connection
between mass of conical singularity and off-shell $\beta$-function
is given by
\begin{equation}
M_{cs}=\frac{\beta}{4\pi}=-M_{pp},
\end{equation}
where $M_{pp}$ is the mass of point particle at the event horizon.
In general, the mass of a conical singularity is negative.

We find from Fig. 3 that the off-shell $\beta$-functions are
negative for $T>T_m$, while these are positive for $T<T_m$. We
have positive $\beta$-function between two equilibrium points,
$r_i<r_+<r_f$. Hence these are not regarded as a properly defined
off-shell process of increasing black hole by absorbing radiation.
However, its reverse process of decreasing black hole by emitting
radiation as the Hawking radiation seems to be possible to occur.

\section{AdSRN black hole}

We consider the AdSRN black hole whose metric function  is given
by
\begin{equation} \label{adsrnMF}
U(r)=1-\frac{2M}{r}+\frac{Q^2}{r^2}+\frac{r^2}{l^2},
\end{equation}
where $l$ is the curvature radius of AdS space. Then, the inner
($r_-$) and the outer ($r_+$) horizons are obtained from the
condition of  $U(r_\pm)=0$. In this work we consider the case of
fixed-charge ensemble $Q<Q_c=l/6$ ~\cite{CEJM}. The other case of
the fixed potential $\Phi=Q/r_+$ will have parallel with the fixed
charge case.

For the AdSRN black hole, the relevant thermodynamic quantities
are given by the ADM mass $M$ and Hawking temperature $T_H$
\begin{equation} \label{aas}
M(r_+,Q)=\frac{r_+}{2}\Big(1+\frac{Q^2}{r_+^2}+\frac{r_+^2}{l^2}\Big),~
T_{H}(r_+,Q)= \frac{1}{4\pi}\Big(
\frac{1}{r_+}-\frac{Q^2}{r_+^3}+\frac{3r}{l^2}\Big).
\end{equation}
In the case that the horizon is degenerate, we have an extremal
black hole with $M=M_e$. In general, one has an inequality of
$M>M_e$. Then, using the Eqs. (\ref{aas}), the heat capacity
$C=(dM/dT_{H})_Q$ for fixed charge and Helmholtz free energy $F$
are obtained to be
\begin{eqnarray}\label{aac}
C(r_+,Q)&=& 2\pi r_+^2 \Big[\frac{3r_+^4+l^2(r_+^2-Q^2)}{3r_+^4+l^2(-r_+^2+3Q^2)}\Big], \\
\label{aaf}
F(r_+,Q)&=&E-T_{H}S_{BH}=\frac{1}{4r_+}\Big(r_+^2+3Q^2-\frac{r_+^4}{l^2}\Big)-M_e
\end{eqnarray}
with $E=M-M_e$. In this case, $E$ measures the energy above the
ground state.  Also  the extremal black hole is chosen  as
background for fixed-charge ensemble~\cite{CEJM}.

\begin{figure*}[t!]
   \centering
   \includegraphics{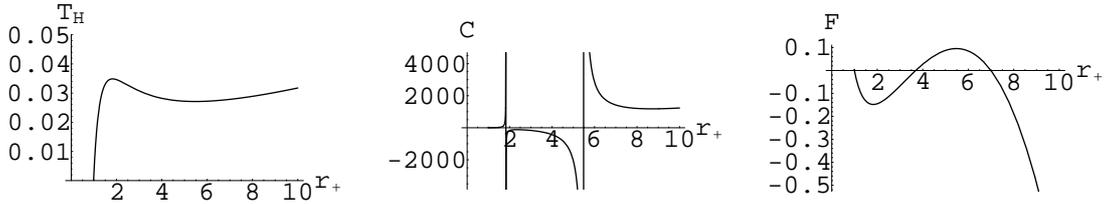}
\caption{Thermodynamic quantities of the AdSRN black hole as
function of horizon radius $r_+$ with fixed $Q=1$ and $l=10$:
temperature $T_H$, heat capacity $C$, and free energy $F$.}
\label{fig4}
\end{figure*}
The global features of thermodynamic quantities are shown in Fig.
4. It seems to be a combination of RN and AdS Schwarzschild black
holes. Here we observe the local minimum $T_H=T_0$ at $r_+=r_{0}$
(feature of AdS Schwarzschild black hole) , in addition to the
extremal temperature $T_H=0$ at $r_+=r_e$ and the maximum value
$T_H=T_m$ at $r_+=r_{m}$ (feature of RN black hole).

 For $r_e<r_+<r_m$, the black hole is locally stable because of $C>0$
while for $r_m<r_+<r_0$, it is locally  unstable ($C<0$). For
$r_+>r_0$, the black hole becomes stable because of $C>0$.
 Here,
we observe that $C=0$ at $r_+=r_e$, and  it blows up at $r_+=r_m,
r_0$. Based on  the local stability, the AdSRN black holes are
split into SBH with $C>0$ being in the region of $r_e<r_+<r_m$,
IBH with $C<0$ in the region of $r_m<r_+<r_0$, and LBH with $C>0$
in the region of $r_+>r_0$.

Importantly, the free energy plays a crucial role  to test the
phase transition. A black hole is   globally stable when $C>0$ and
$F<0$. We observe two extremal points for free energy: the local
minimum $F=F_{min}$ at $r_+=r_{m}$ and the maximum value
$F=F_{max}$ at $r_+=r_{0}$. The free energy  is negative for
$r_e<r_+<r_m$ and it increases  in the region of $r_m<r_+<r_0$.
For $r_+=r_1>r_0$, it is zero and remains negative for $r_+>r_1$.
The temperature $T=T_1$ determined from $F=0$ at $r_+=r_1$ plays a
role of the critical temperature in HP2.

\begin{figure*}[t!]
   \centering
   \includegraphics{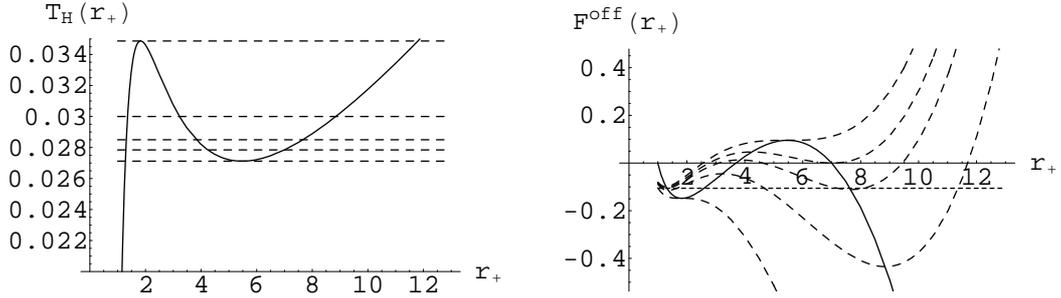}
\caption{Graphs for temperature and free energy expressed as
functions of $r_+$ with $l=10$. In the left panel, the solid curve
represents the Hawking temperature $T_H(r_+,Q=1)$, while five
dashed lines denote
$T=T_m=0.0348,~0.03,~T_2=0.0285,T_1=0.0278,T_0=0.0271$ from top to
bottom. In the right panel, the solid curve represents the
on-shell free energy $F(r_+,Q=1)$, while five dashed curves denote
the off-shell free energy $F^{off}(r_+,Q=1,T)$ for
$T=T_0,T_1,T_2,0.03,T_m$ from top to bottom.} \label{fig5}
\end{figure*}

\begin{figure*}[t!]
   \centering
   \includegraphics{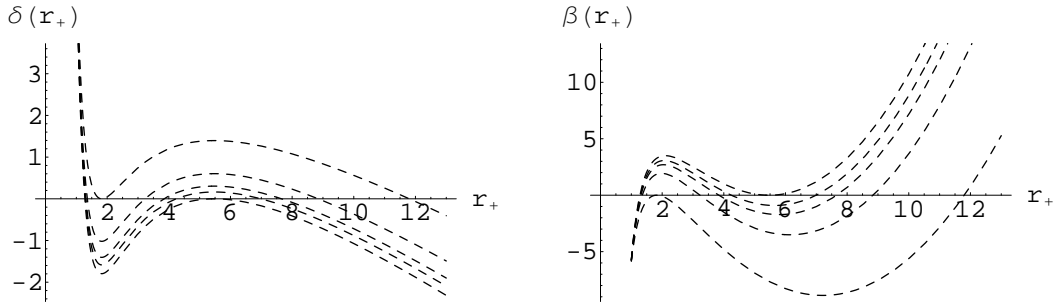}
\caption{Graphs for deficit angle and $\beta$-function. Five
dashed curves denote $\delta(r_+,Q=1,T)$ for
$T=T_m,0.03,T_2,T_1,T_0$ from top to bottom. In the right panel,
five dashed curves denote $\beta(r_+,Q=1,T)$ for
$T=T_0,T_1,T_2,0.03,T_m$ from top to bottom.} \label{fig6}
\end{figure*}

In order to investigate whether the HP1 is possible to occur in
the AdSRN black holes, we study the off-shell process of the
growth of a black hole. For this purpose, we introduce the
off-shell free energy
\begin{equation} \label{offf}
F^{off}(r_+,Q,T)=E-TS_{BH}
\end{equation}
with the  temperature  $T$ of heat reservoir. The  critical
temperature $T=T_2$  is derived from the condition of
$F^{off}(r_s,1,T_2)=F^{off}(r_l,1,T_2)$, which means that the free
energy of SBH at  $r_+=r_s$ is equal to free energy of LBH at
$r_+=r_l$. This case is marked as the horizontal line in the right
panel of Fig. 5.
 Then we
expect to discuss a HP1 between SBH and LBH through IBH for
$T_m=0.0348<T<T_2=0.0285$. As is shown in Fig.5, we choose
$T=0.03$ for a transition temperature of the HP1. This transition
consists of two processes: SBH with $C>0\to$ IBH with $C<0 \to$
LBH with $C>0$. We have   global stabilities for SBH and LBH
because of their free energies are positive. However, the IBH
remains unstable and it may play a role of a mediator for HP1 as
an unstable small black hole in  HP2.

Let us study the HP1 further by introducing the deficit angle
$\delta(r_+,Q,T)$ and the off-shell $\beta$-function
$\beta(r_+,Q,T)$ which are the same forms as in Eqs.(\ref{delta})
and (\ref{beta})  but different temperature $T_H$. From  Fig. 6,
we find that the HP1 is not possible to occur. The reason is that
for the process of SBH$\to$ IBH, the off-shell (non-equilibrium)
process is not well-defined because the negative deficit angle and
positive off-shell $\beta$-function exist for $T=0.03$.  Hence, it
seems that the HP1 is hard to occur in the AdSRN black holes.

\section{ Discussions}
\begin{table}
\centering
\begin{tabular}{|c||c|c||c|c|}
\hline
 &\multicolumn{2}{c||}{APT ($\to$)}&\multicolumn{2}{c|}{EP($\leftarrow$)}\\
\cline{2-5}
 &starting point& ending point & starting point & ending point\\
\hline\hline
    $r_+$   & $r_+=r_i$(NEBH) &$r_+=r_f$(NOBH)& $r_+=r_f$ &$r_+=r_i$\\ \hline
  $C$ &$+$ & $-$&$-$&+ \\ \hline
  $F$ & $ -$ & $+$  &  $+$ &$-$ \\ \hline
 status  & stable & unstable & unstable  & stable \\
\hline
\end{tabular} \caption{Summary of the assumed phase transition (APT) and
its reverse transition as Evaporating process (EP) for the RN
black hole.}
\end{table}

\begin{table}
\centering
\begin{tabular}{|c||c|c|c|}
\hline
 &starting point& intermediate point & ending point \\
\hline\hline
    $r_+$   & $r_+=r_s$(SBH) &$r_+=r_i$(IBH)&$r_+=r_l$(LBH) \\ \hline
  $C$ &$+$ & $-$&$+$ \\ \hline
  $F$ & $ -$ & $+$  & $-$ \\ \hline
 status  & stable & unstable & stable   \\
\hline
\end{tabular} \caption{Summary of the assumed Hawking-Page transition (HP1:$\to$) for the AdSRN black hole.}
\end{table}

We discuss the phase transition between EBH and NOBH in the RN
black hole. For this purpose, we may consider this transition as
the
 $T \to 0$ limiting case of the assumed phase transition between the NEBH and NOBH.
 Our study was based on the on-shell observations of temperature,
heat capacity and  free energy as well as the off-shell
observations of generalized (off-shell) free energy, deficit angle
and $\beta$-function.  In general, the on-shell thermodynamics
describes relationships among thermal equilibria and not the
transitions between equilibria. The off-shell thermodynamics is
suitable  for the description of  transitions between thermal
equilibria. As is summarized in Table 1, the NEBH is a globally
stable object, whereas the NOBH is an unstable object.
 The $\leftarrow$ transition  from NOBH
to NEBH is likely allowed as an evaporating process~\cite{Pavon},
while the assumed phase transition (NEBH$\to$NOBH) is unlikely
allowed for $T< T_m$~\cite{myung2}. The main reason comes from the
fact that for the latter, the off-shell (non-equilibrium) process
is not well-defined because of the negative deficit angle and
positive off-shell $\beta$-function. As is shown in Fig. 2, this
arises because of the presence of temperature barrier.
Furthermore, an evaporating process from NEBH to EBH is possible
to occur as was discussed in Ref.\cite{FNN}.

In connection with HP1, we have a correspondence of NEBH
$\leftrightarrow$ SBH and NOBH $\leftrightarrow$ IBH. The process
of SBH$\to$IBH represents the feature of RN black hole, while the
process of IBH$\to$LBH denotes the feature of AdS Schwarzschild
black hole.  Thus, one expects that for $T<T_2$, the SBH is more
favorable than the LBH, while for $T>T_2$, the LBH is more
favorable than the SBH.
 However,  the HP1 of SBH$\to$IBH$\to$LBH is hard to
occur because   SBH$\to$IBH (NEBH$\to$NOBH) is included as the
first process in the AdSRN black holes. The presence of
temperature barrier in Fig. 4 supports this argument. Further, its
reverse transition is not allowed because both starting and ending
points are globally stable (see Table 2).

Finally, we propose that NEBH$\nrightarrow$NOBH and
NEBH$\leftarrow$NOBH for the RN black holes, while
SBH$\nrightarrow$LBH and SBH$\nleftarrow$LBH for the AdSRN black
holes.

 We conclude that an evaporating
process from non-extremal black hole (NOBH) to extremal one (EBH)
is possible to occur, but  its reverse process  is not possible to
occur. However,  considering  the quantum tunnelling process
through temperature barrier, one may expect to have NEBH$\to$NOBH
 for the RN black holes and SBH$\to$LBH
for the AdSRN black holes.

\section*{Acknowledgement}
The author thanks Young-Jai Park and Yong-Wan Kim for helpful
discussions. This work was supported by the Korea Research
Foundation (KRF-2006-311-C00249) funded by the Korea Government
(MOEHRD).

\end{document}